\documentclass[aps,prd,reprint,amsmath,amssymb,showpacs,superscriptaddress,nofootinbib]{revtex4-1}
\usepackage{epsfig,slashed,xcolor,bm,natbib,comment,enumitem,ulem,cancel}
\usepackage[cal=boondox]{mathalfa}
\def\vect#1{\mbox{\boldmath $#1$}}
\def\hvect#1{{\hat{\mbox{\boldmath $#1$}}}}

\def\nd#1#2{{\frac{d #1}{d #2}}}

\def\tfrac#1#2{{\textstyle\frac{#1}{#2}}}

\newcommand{\vpsi}{\varphi}

\newcommand{\comma}{, }
\newcommand{\be}{\begin{equation}}
\newcommand{\ee}{\end{equation}}
\newcommand{\bea}{\begin{eqnarray}}
\newcommand{\eea}{\end{eqnarray}}
\newcommand{\myDel}[1]{{\color{red}\ifmmode\cancel{#1}\else\sout{#1}\fi}}

\begin{document}
\title{Note on the absence of the second clock effect in Weyl gauge
  theories of gravity} 
\author{M.P.~\surname{Hobson}}
\email{mph@mrao.cam.ac.uk} 
\affiliation{Astrophysics Group, Cavendish
  Laboratory, JJ Thomson Avenue, Cambridge CB3 0HE\comma UK}
\author{A.N.~\surname{Lasenby}} \email{a.n.lasenby@mrao.cam.ac.uk}
\affiliation{Astrophysics Group, Cavendish Laboratory, JJ Thomson
  Avenue, Cambridge CB3 0HE\comma UK} \affiliation{Kavli Institute for
  Cosmology, Madingley Road, Cambridge CB3 0HA, UK}
\date{Received 21 December 2021; accepted 3 January 2022}

\begin{abstract}
We reconsider the status of the so-called second clock effect in Weyl
gauge theories of gravity, which are invariant both under local
Poincar\'e transformations and local changes of scale. In particular,
we revisit and extend our previous demonstration that the second clock
effect does not occur in such theories, in order both to clarify our
argument and confirm our original findings in response to
recent counterclaims in the literature.
\end{abstract}

\maketitle

\enlargethispage{0.8cm}
Just three years after Einstein proposed his theory of general relativity, Weyl
applied the principle of relativity not only to the choice of
reference frames, but also to the choice of local standards of length
(or `gauge') \cite{Weyl18}. To achieve invariance under the latter, Weyl
introduced an additional `compensating' vector field, that we shall
denote by $B_\mu$, which he attempted to interpret as the
electromagnetic 4-potential in a unified theory of gravity and
electromagnetism. Although elegant and beautiful, Weyl's theory
was soon recognized as being unable to accommodate well-known properties of
electromagnetism, which was only later realised to be
related to localisation of invariance under change of
quantum-mechanical phase \cite{Weyl31}. Nonetheless, $B_\mu$ may instead be
interpreted as mediating an additional {\it gravitational}
interaction, within a locally scale-invariant theory of gravity. Even
in this context, however, Weyl's theory remains subject to an
objection first published by Einstein as an addendum to Weyl's
original paper, namely that it predicts a so-called `second clock
effect' (SCE), which is not experimentally observed.
This is in addition to the usual `first clock effect',
which also occurs both in special and general relativity and is
experimentally verified to high precision. 

In particular, Einstein considered two identical clocks that are
initially synchronised, coincident and at relative rest, but then
follow different timelike worldlines in spacetime before coming back
together. He claimed that Weyl's theory predicts the clocks will
`tick' at {\it different rates} even after they are reunited, if the
field strength $H_{\mu\nu} \equiv 2\partial_{[\mu} B_{\nu]}$ of the
potential does not vanish on any open surface bounded by the two clock
worldlines during their separation. This has the physical consequence
of precluding the existence of sharp spectral lines, since the rate of
atomic clocks would depend on their past history.

The SCE was a source of considerable controversy between Weyl,
Einstein, Eddington and Pauli, amongst others
\cite{Pauli19,Eddington20,Eddington24,Schulman97,Goenner04} and has
remained a matter of much debate ever since
\cite{Perlick87,Cheng88,Novello92,Wheeler98,Spencer11,Wheeler13,deCesare17,Wheeler18,Avalos18,Lobo18,Scholz18}. In
\cite{Hobson20}, we entered this debate by investigating whether the
SCE occurs in Weyl gauge theories of gravity (WGTs)
\cite{Dirac73,Bregman73,Charap74,Kasuya75,Harnad76,
  Blagojevic02}, which we have considered (along with some
extensions) as promising candidate modified gravity theories in a
number of contexts
\cite{eWGTpaper,Barker20a,Barker20b,Lin21,Barker21}.

As we pointed out in \cite{Hobson20}, two complementary approaches to
investigating the SCE have been considered previously. First, the
original arguments for the SCE were based on the fact that in a Weyl
spacetime, the `norms' of vectors that are parallel transported
according to the standard $W_4$ covariant derivative $\nabla_\mu$ will
change in a manner that depends on the path taken. Assuming that the
norm of a timelike vector that is parallel transported along a
timelike worldline can represent the `tick' rate of a clock, this
therefore leads to a SCE unless the Weyl potential can be expressed as
the gradient of some smooth scalar field $B_\mu = \partial_\mu\vpsi$,
such that its field strength $H_{\mu\nu}$ vanishes identically.  The
association of the norm of a timelike vector with a clock rate is not
trivial, however, particularly given that length is not a well-defined
concept in Weyl's spacetime. Thus, a different approach has been
proposed where one instead defines an alternative notion of proper
time along (timelike) worldlines, which generalises the concept of
proper time used in Riemannian spacetimes and is claimed to be well
defined in Weyl spacetime \cite{Ehlers12, Perlick87, Avalos18,
  Lobo18}. If one then reconsiders the two-clock thought experiment
mentioned above, and computes this particular elapsed proper time as
measured by each clock between their reunion and some subsequent
event, the two measurements differ and so it is again usually
concluded that a Weyl spacetime exhibits a SCE.

Contrary to this prevailing view, we demonstrated in \cite{Hobson20}
that the SCE does not occur in WGTs (with or without torsion, which is
irrelevant to the SCE). In particular, in the first part of our
argument we showed that the geometric interpretation of WGTs leads to
the identification of the Weyl covariant derivative $\nabla^\ast_\mu =
\nabla_\mu + wB_\mu$ as the natural derivative operator, which differs
from the covariant derivative $\nabla_\mu$ usually assumed in
Weyl(--Cartan) spacetimes when applied to quantities having non-zero
scaling dimension (or Weyl weight) $w$.  This is especially important
when differentiating the tangent vector $u^\mu(\lambda) =
dx^\mu/d\lambda$ along an observer's worldline, which we showed must
have Weyl weight $w=-1$, rather than being invariant ($w=0$) as is
usually assumed. One then finds directly that the norm of a
parallel-transported tangent vector does not depend on the path taken,
hence addressing the original argument for suggesting the presence of
the SCE. To address the complementary approach framed in terms of
elapsed proper times for the two clocks considered above, we pointed
out in the second part of our argument that, since Einstein's
objection to Weyl's theory is based on the observation of sharp
spectral lines, one requires the presence of matter fields to
represent atoms, observers and clocks; it is thus meaningless to
consider the SCE in an empty Weyl(--Cartan) geometry.  Moreover, such
`ordinary' matter is most appropriately represented by a massive Dirac
field, but in order to obey local Weyl invariance this field must
acquire a mass dynamically through the introduction of a scalar
compensator field $\phi$, which we showed is key to defining an interval of
proper time as measured by a clock along an observer's
worldline.\footnote{One should note that the compensator scalar
  field $\phi$ is, in general, totally unrelated to the scalar field
$\vpsi$ that defines the Weyl vector $B_\mu = \partial_\mu\vpsi$
in an integrable Weyl spacetime.} On
taking these considerations into account, we showed that the two
clocks experience the same proper time between their reunion and any
subsequent event, so that no SCE occurs, even when the Weyl potential
is not pure gauge.

It has recently been claimed, however, that our previous analysis
contains a flaw and that the SCE does indeed occur in WGTs
\cite{Quiros21}. This difference of opinion results partly from a lack
of clarity in our original presentation, but more significantly 
from the analysis in \cite{Quiros21} not properly taking into
  account the requirement in the second part of our argument for the
  introduction of the scalar compensator field.  As we will now show,
when one carefully considers this aspect of our full
argument, one finds that the SCE does not, in fact, occur, thereby
confirming our original findings.

The outline of the first part of our argument in \cite{Hobson20} is as
follows. The geometric interpretation of WGTs identifies the (inverse)
translational gauge field as the vierbein components ${e^a}_\mu$,
which have Weyl weight $w=1$ and relate the orthonormal tetrad frame
vectors $\hvect{e}_a(x)$ and the coordinate frame vectors
$\vect{e}_\mu(x)$ at any point $x$ in a Weyl--Cartan spacetime.  The
vectors $\hvect{e}_a(x)$ constitute a local Lorentz frame at each
point, which defines a family of ideal observers whose worldlines are
the integral curves of the timelike unit vector field
$\hat{\mathbf{e}}_0$. For some test particle (or other observer)
moving along some timelike worldline $\mathcal{C}$ given by
$x^\mu=x^\mu(\lambda)$, where $\lambda$ is some arbitrary parameter,
the components of the tangent vector to this worldline, as measured by
one of the above observers, will be $u^a(\lambda) = {e^a}_\mu
u^\mu(\lambda)$, which are physically observable quantities in WGTs
and so should be invariant ($w=0$) under the (simultaneous) Weyl scale
gauge transformations $g_{\mu\nu} \to e^{2\rho} g_{\mu\nu}$ and $B_\mu
\to B_\mu -\,\partial_\mu\rho$, where $\rho=\rho(x)$ is an arbitrary
scalar function. Since the vierbein ${e^a}_\mu$ has weight $w=1$, the
weight of the components $u^\mu(\lambda)$ must thus be
$w=-1$.\footnote{As we also discuss in \cite{Hobson20}, one may reach
  the same conclusion by demanding that the physical distance, as
  opposed to the coordinate distance, along the curve $\mathcal{C}$ is
  traced out at the same rate before and after a Weyl scale gauge
  transformation.} The length of the tangent vector $u^a(\lambda)$ is
then invariant under Weyl scale gauge transformations. Moreover, we
also pointed out in \cite{Hobson20} that one should work exclusively
in terms of the Weyl covariant derivative $\nabla^\ast_\mu =
\nabla_\mu + wB_\mu$, which is fully covariant and has the important property
$\nabla^\ast_\mu g_{\rho\sigma}=0$. Thus, for example, parallel
transport along a worldline $x^\mu = x^\mu(\lambda)$ of the coordinate
components $v^\mu$ of any vector should be  defined by
\be
\frac{D^\ast v^\mu}{D\lambda} \equiv u^\sigma\nabla^\ast_\sigma v^\mu = 0,
\label{eqn:parat}
\ee
rather than by $Dv^\mu/D\lambda \equiv u^\sigma\nabla_\sigma v^\mu = 0$, as
had been assumed by previous authors.
One may then immediately show that for
any two vectors {\it of weight} $w=-1$ parallel transported according to
(\ref{eqn:parat}), one has
\bea
\nd{}{\lambda}[g_{\mu\nu}v^\mu(\lambda)z^\nu(\lambda)] 
&=& \frac{D^\ast}{D\lambda}[g_{\mu\nu}v^\mu(\lambda)z^\nu(\lambda)],\nonumber\\
&=& (u^\sigma(\lambda)\nabla^\ast_\sigma g_{\mu\nu}) v^\mu(\lambda)
z^\nu(\lambda) =0,\phantom{AAa}
\label{eqn:dpcons}
\eea
which is equation (49) in \cite{Hobson20}.  Hence, by setting
$v^\mu=z^\mu$ and considering parallel transport around a closed curve
${\cal C}$, the norm of any vector {\it of weight $w=-1$} is unchanged
on completing a loop. Thus, the norm of the parallel-transported
tangent vector $u^\mu = dx^\mu/d\lambda$ itself is unchanged, and so
the original basis for suggesting the existence of a SCE is removed,
as it is {\it this} norm that was taken to represent the clock rate.

Since we were concerned in \cite{Hobson20} exclusively with
(coordinate) vectors of weight $w=-1$, however, we did not clarify --
as correctly pointed out in \cite{Quiros21} -- that (\ref{eqn:dpcons})
holds {\it only} if the scalar $g_{\mu\nu}v^\mu z^\mu$ has weight
$w=0$ (recalling that $w(g_{\mu\nu}) = 2$), although the first
equality in (\ref{eqn:dpcons}) implicitly contains this
assumption. More generally, if the weights of the vectors $v^\mu$ and
$z^\mu$ are $w_v$ and $w_z$, respectively, and denoting
$\vect{v}\cdot\vect{z} \equiv g_{\mu\nu}v^\mu(\lambda)z^\nu(\lambda)$
and $\vect{u}\cdot\vect{B}=u^\mu(\lambda)B_\mu$ for brevity, one
immediately has
\bea
\nd{}{\lambda}(\vect{v}\cdot\vect{z})
&=& \left[\frac{D^\ast}{D\lambda}-(2+w_v+w_z)(\vect{u}\cdot\vect{B})\right](\vect{v}\cdot\vect{z}),\nonumber\\
&=& -(2+w_v+w_z)(\vect{u}\cdot\vect{B})(\vect{v}\cdot\vect{z}),\phantom{AAa}
\label{eqn:dpconsw}
\eea
which is equivalent to equation (180) in \cite{Quiros21}, although the
latter is derived at somewhat greater length. As noted in
\cite{Quiros21}, equation (\ref{eqn:dpconsw}) implies that the inner
product of two non-orthogonal parallel-transported vectors is path
independent {\it only if} $w_v + w_z = -2$ (unless the Weyl vector is
pure gauge $B_\mu = \partial_\mu\vpsi$). It is then further claimed in
\cite{Quiros21}, however, that this implies that the SCE does occur,
since one may take $v^\mu$ and $z^\mu$ to coincide with the 4-momentum
$p^\mu$ of a particle, which has weight $w(p^\mu)=-2$, so that the
mass (squared) $m^2 = g_{\mu\nu}p^\mu p^\nu$ of an atom, and hence the
frequencies of its spectral lines, will depend on its past history.
We now show that this further claim is unjustified on account of the
requirement in the second part of our original argument presented in
\cite{Hobson20} that one must introduce a scalar compensator field;
this results in the 4-momentum of a particle {\it not} being parallel
transported along its worldline, even if it is in free-fall
(i.e.\ moving only under gravity), so that
(\ref{eqn:dpconsw}) does not apply.

As we noted in \cite{Hobson20}, in order for WGTs to include
`ordinary' matter necessary for atoms, observers and clocks, which is
usually modelled by a Dirac field, one must introduce a scalar
compensator field $\phi$ with Weyl weight $w=-1$ and make the
replacement $m\bar{\psi}\psi \to \mathcal{m}\phi\bar{\psi}\psi$ in the
Dirac action, where $\mathcal{m}$ is a dimensionless parameter but
$\mathcal{m}\phi$ has the dimensions of mass in natural units. In this
way, particle masses are not fundamental, which is forbidden by
scale-invariance, but instead arise dynamically. Since the functioning
of any form of practical atomic clock is based on the spacing of the
energy levels in atoms, let us consider an atom with dimensionless
parameter $\mathcal{m}$ that traces out some timelike worldline
$x^\mu=x^\mu(\lambda)$ in spacetime.

The key physical question to address is how the 4-velocity $u^\mu$ and
4-momentum $p^\mu$ of the atom evolve along its worldline. The
definition in (\ref{eqn:parat}) for parallel transport is fully
covariant and, in particular, when applied to $u^\mu$ or $p^\mu$ it is
analogous to the familiar expression for geodesic motion of an atom in
free fall in general relativity (or some other non-scale-invariant
gravity theory), where the particle has a fixed kinematic mass $m$.
Indeed, this is {\it why} one considers parallel transport of the
4-velocity or 4-momentum (or even the spin 4-vector) along the
worldline of a particle in free fall in such theories.  Even in
general relativity, however, for a particle not in free fall, the
4-velocity and 4-momentum are {\it not} parallel transported along the
particle worldline, but instead their rates of change at any point are
proportional to the particle 4-acceleration there. Thus, in general
relativity, the concept of parallel transport of the 4-velocity or
4-momentum of a particle is relevant {\it only} if it is in free fall
and hence following a timelike {\it geodesic} worldline.

As we now show, in scale-invariant gravity theories, the dynamics of
an atom with a dynamically-generated mass is such that {\it neither}
its 4-velocity {\it nor} 4-momentum is parallel transported according
to (\ref{eqn:parat}) along its worldline, {\it even} if the atom is in
free fall. As we have described
previously in \cite{eWGTpaper,Hobson21}, assuming ordinary matter to
be represented by a Dirac field, one may construct an appropriate
action for a spin-$\tfrac{1}{2}$ point particle and then transition to
the full classical approximation in which the particle spin is
neglected. In the presence of a Yukawa coupling to a scalar
compensator field $\phi$, this yields
\begin{equation}
S_{\rm p} = - \int d\lambda\, [p_a u^{a}-\tfrac{1}{2}e(p_a p^a - \mathcal{m}^2\phi^2)],
\label{WGTppaction}
\end{equation}
where the dynamical quantities are the tetrad components of the
particle 4-momentum $p_a(\lambda) = {e_a}^\mu p_\mu(\lambda)$ and
4-velocity $u^a(\lambda) = {e^a}_\mu dx^\mu(\lambda)/d\lambda$, and
the einbein $e(\lambda)$ along the worldline $x^\mu(\lambda)$.  On
varying the action with respect to the dynamical variables $p_a$,
$x^\mu$ and $e$, one finds that the equations of motion for the
particle may
then be written in the coordinate frame as
\bea
u^\mu & = & ep^\mu, \label{WGTppeom1}\\
u^\sigma\nabla^\ast_\sigma p_\mu  & = &
e\mathcal{m}^2 \phi\,\nabla^\ast_\mu\phi + u^\sigma T^\ast_{\sigma\mu\nu}p^\nu, \label{WGTpppdot}\\
p^2 &=& \mathcal{m}^2\phi^2,\label{WGTppeom3}
\eea
where $p^2 \equiv p_a p^a = p_\mu p^\mu$ and $T^\ast_{\sigma\mu\nu}$
is the Weyl--Cartan torsion. In order that $u^au_a = u^\mu u_\mu = 1$
for a massive particle, the einbein must take the form
$e=1/(\mathcal{m}\phi)$, and so  $p^\mu = \mathcal{m}\phi u^\mu$, as
expected.  In this case, the weights of the quantities appearing in
(\ref{WGTppaction}) are $w(p_a)=-1$, $w(u^a)=0$, $w(e)=1$,
$w(\lambda)=1$ and $w(\vpsi)=-1$, so that the action is indeed
scale-invariant. As we show below, the dynamics of the particle are
consistent with the length of the 4-velocity remaining equal to unity
along the particle worldline; this agrees with our finding in
\cite{Hobson20} that one may always obtain a parameterisation
$\xi=\xi(\lambda)$ for which the length of the tangent vector remains
equal to unity along its worldline (and so $u^\mu(\xi)=dx^\mu/d\xi$
may be interpreted as the particle 4-velocity). Indeed, to harmonise
our notation with that used in \cite{Hobson20}, we will take the
opportunity at this point to relabel the parameter along the worldline
as $\xi$.

From (\ref{WGTppeom1}) and (\ref{WGTpppdot}), one finds that,
respectively, the 4-velocity and 4-momentum of a particle in free fall
are transported along its worldline according to
\bea u^\sigma\nabla^\ast_\sigma u_\mu & = & (\delta_\mu^\sigma - u_\mu
u^\sigma)\nabla^\ast_\sigma\ln\phi + u^\sigma
T^\ast_{\sigma\mu\nu}u^\nu \label{eqn:utrans}\\ u^\sigma\nabla^\ast_\sigma
p_\mu & = & \mathcal{m} \nabla^\ast_\mu\phi + u^\sigma
T^\ast_{\sigma\mu\nu}p^\nu. \label{eqn:ptrans} 
\eea
Thus, it is only when
$\nabla_\mu^\ast\phi=\partial_\mu\phi-B_\mu\phi=0$ (such that the Weyl
vector is pure gauge $B_\mu = \partial_\mu\ln\phi$) and the torsion
vanishes that the 4-velocity and 4-momentum are parallel transported
according to (\ref{eqn:parat}) along the particle's worldline. In
particular, (\ref{eqn:utrans}) shows that a particle moving only under
gravity has, in general, a {\it non-zero} covariant 4-acceleration $a^{\ast\mu}
\equiv u^\sigma \nabla^\ast_\sigma u^\mu$, as defined in
\cite{Hobson20}, and equivalently experiences a {\it non-zero}
covariant 4-force $f^{\ast\mu} \equiv u^\sigma \nabla^\ast_\sigma
p^\mu$. Nonetheless, as noted above, $u^2=u_\mu u^\mu$ (which has
weight $w=0$) does {\it not} change along the worldline, since
\be
\nd{u^2}{\xi}  =  u^\sigma\nabla^\ast_\sigma u^2 
 =  2 u_\mu u^\sigma\nabla^\ast_\sigma u^\mu = 0,
\label{eq:ua0}
\ee
where the final equality holds on using (\ref{eqn:utrans}) and noting
that $T^\ast_{\sigma\mu\nu} = -T^\ast_{\sigma\nu\mu}$. By contrast,
the evolution of $p^2$ (which has weight $w=-2$) along the worldline
is
\be \nd{p^2}{\xi}  =  (u^\sigma\nabla^\ast_\sigma +
2u^\sigma B_\sigma) p^2 = 2 p_\mu
u^\sigma\nabla^\ast_\sigma p^\mu + 2u^\sigma B_\sigma p^2.  
\ee 
On using (\ref{eqn:ptrans}) to substitute for $u^\sigma\nabla^\ast_\sigma
p^\mu$ and recalling that $w(\phi) = -1$,
one obtains
\be
\nd{p^2}{\xi} =
2\mathcal{m}^2\phi\left(\nd{\phi}{\xi}-u^\sigma B_\sigma \phi\right)
+ 2u^\sigma B_\sigma p^2 = 2\mathcal{m}^2\phi\nd{\phi}{\xi}.
\ee
Hence, one has $d\ln p/d\xi = d\ln\phi/d\xi$, which is {\it
  independent} of the (free-fall) path taken and expresses merely the
fact that $p(x) \propto \phi(x)$, as expected.

Thus, one can infer that {\it no} SCE occurs as a result of the
transport of the 4-velocity or 4-momentum along the worldline of a
particle in free fall, thereby confirming our original findings in
\cite{Hobson20}. It is worthwhile illustrating this point in a
  specific scenario. One may, for example, consider an atom following
  a closed free-fall orbit in a static spacetime with a non-zero field
  strength $H_{\mu\nu}=2\partial_{[\mu}B_{\nu]}$ of the Weyl
  potential. After completing each orbit, the atom will return to the
  same point with the norms of both its 4-velocity and 4-momentum
  having their original values there, hence demonstrating that no SCE
  occurs.\footnote{If the atom has spin, then a fuller
    treatment would be necessary, but the outcome is likely still to
    be as above if one considers a cloud of unaligned atoms, which is
    probably a more realistic representation of an atomic clock.}
In particular, the need to include the effect of the $\phi$
  field leads to evolution of the 4-velocity and 4-momentum along the
  worldline that differs from parallel transport, which shows that the
  latter {\it cannot} be a general prescription. This is, of course,
  at variance with what other authors have typically assumed, but is
  nevertheless a strong conclusion of this work. Moreover, since we
  have shown that this leads to the {\it absence} of a second clock
  effect, the objections presented in \cite{Quiros21} to our previous
  claims regarding the SCE are certainly not valid for the case we
  have discussed above.

The question does remain, however, as to how one may extend our
  analysis to a particle {\it not} in free fall. When considering a
  real physical atom, this is likely to be a complicated issue that we
  do not wish to address in detail here.  Nonetheless, for an ideal
  clock one may employ an approach analogous to that used
  in general relativity. As shown in \cite{Hobson20}, in a
  Weyl(--Cartan) spacetime one may always
  find a parameterisation $\xi=\xi(\lambda)$ for which the length of
  the tangent vector remains equal to unity along {\it any} timelike
  worldline, such that
\be
\nd{u^2}{\xi}  =  u^\sigma\nabla^\ast_\sigma u^2 
 =  2 u_\mu a^{\ast\mu}= 0,
\label{eq:genua0}
\ee
where the covariant 4-acceleration $a^{\ast\mu} \equiv u^\sigma
\nabla^\ast_\sigma u^\mu$ may have a more general form than in
(\ref{eqn:utrans}). In particular, one may write $a^{\ast\mu} =
a^{\ast\mu}_{\rm g} + a^{\ast\mu}_{\rm p}$, where the `gravitational'
4-acceleration $a^{\ast\mu}_{\rm g}$ represents the terms on the
right-hand side of (\ref{eqn:utrans}) and $a^{\ast\mu}_{\rm p}$ is an
additional `peculiar' 4-acceleration. Since (\ref{eq:ua0}) shows that
$u_\mu a^{\ast\mu}_{\rm g}=0$, one also has $u_\mu a^{\ast\mu}_{\rm
  p}=0$.  Thus (\ref{eq:genua0}) is essentially a generalisation of
the clock hypothesis, whereby an infinitesimal time interval measured
by an ideal clock with some peculiar acceleration equals that
measured in a momentarily comoving inertial frame (although this may
well not hold for a physical atom). It is straightforward to show that
the 4-force then has the analogous form $f^{\ast\mu} \equiv u^\sigma
\nabla^\ast_\sigma p^\mu = f^{\ast\mu}_{\rm g} + f^{\ast\mu}_{\rm p}$,
where $f^{\ast\mu}_{\rm g}$ represents the terms on the right-hand
side of (\ref{eqn:ptrans}) and $f^{\ast\mu}_{\rm p} =
\mathcal{m}\phi\,a^{\ast\mu}_{\rm p}$. Moreover, one therefore finds
that along {\it any} timelike worldline,
\bea \nd{p^2}{\xi} & = & (u^\sigma\nabla^\ast_\sigma + 2u^\sigma
B_\sigma) p^2\nonumber \\ &=& 2\mathcal{m}^2\phi\nd{\phi}{\xi} + 2
p_\mu f^{\ast\mu}_{\rm p} = 2\mathcal{m}^2\phi\nd{\phi}{\xi}, 
\label{eq:gendp2}
\eea
where in the last equality we have used the result $p_\mu
f^{\ast\mu}_{\rm p} = \mathcal{m}^2\phi^2\,u_\mu a^{\ast\mu}_{\rm
  p}=0$.  Thus (\ref{eq:gendp2}) may be considered as a generalisation
of the pure 4-force hypothesis, whereby the rest mass of a particle is
unchanged by the action of any `peculiar' 4-force. From (\ref{eq:genua0}) and
(\ref{eq:gendp2}), one then infers that no SCE
occurs along {\it any} timelike worldline.

Before concluding, we take the opportunity here also to clarify our
argument in \cite{Hobson20} regarding the definition of an
appropriately invariant proper time for a particle moving along any
timelike worldline. Since $d/d\xi$ has weight $w=-1$, the parameter $\xi$
cannot be interpreted as the proper time.  To resolve this issue, one
must again take account of fact that the masses of `ordinary' matter
particles are not fundamental in scale-invariant theories, but instead
arise dynamically through interaction of the Dirac field with the
scalar field $\phi$, such that the particle mass is given by $m =
\mathcal{m}\phi$. In particular, the functioning of an atomic clock is
based on the spacing of the energy levels in atoms, which is
characterised by the Rydberg energy $E_{\rm R} =
\tfrac{1}{2}\mathcal{m}\phi\alpha^2$ (in natural units), where
$\alpha$ is the (dimensionless) fine structure constant, and so in
general varies with spacetime position according to the value of
$\phi$. The quantity $E_{\rm R}$ is defined as the projection of the
4-momentum $p^\mu$ of a photon emitted in a free to ground-state
electronic transition onto the timelike basis vector
$\hat{\mathbf{e}}_0$ of the atom's local Lorentz frame, such that
$E_{\rm R} = g_{\mu\nu} p^\mu dx^\nu/d\xi$.  Therefore, in a small
parameter interval $d\xi$, the number of cycles traversed in the
photon wave train is $dN \propto E_{\rm R}\,d\xi \propto \phi\,d\xi$,
which is invariant under a Weyl gauge transformation, since $\phi$ and
$d\xi$ have weights $w=-1$ and $w=1$, respectively. This corresponds
to the invariance of phase for all observers. A proper time interval
$d\tau$ in the atom's rest frame is {\it defined}, however, as the
duration of a given number of cycles, and so $d\tau \propto
\phi\,d\xi$, where one can take the constant of proportionality to
equal unity, without loss of generality. Hence the proper time
interval measured by an atomic clock between two events corresponding
to the parameter values $\xi_0$ and $\xi$ along the worldline is given
simply by
%
$\Delta\tau(\xi) = \int_{\xi_0}^\xi \phi\,d\xi'$,
%
which is invariant under Weyl gauge transformations, as required for a
physically observable quantity. Indeed, the photon energy measured by
an observer comoving with the atomic clock (which might more
reasonably be called the Rydberg energy) is proportional to the
angular frequency of the photon as measured in terms of the proper
time $\tau$, which is itself invariant under Weyl gauge
transformations, as it should be, and given by $E_{\rm
  R}/\phi=\tfrac{1}{2}\mathcal{m}\alpha^2$. 

One may also arrive at the identification $d\tau = \phi\,d\xi$
for the proper time by a more general route that in addition yields a form
for the particle action that coincides with heuristic expectations.
Varying the particle action (\ref{WGTppaction}) with respect to $p_a$
yields (\ref{WGTppeom1}), which may then be substituted back into 
(\ref{WGTppaction}) to obtain
\be
S_{\rm p} = -\tfrac{1}{2}\int d\lambda\,\left(\frac{1}{e}u^2+e\mathcal{m}^2\phi^2\right).
\label{ppaction2}
\ee
Now varying this action with respect to the einbein $e$ yields
$e=\sqrt{u^2}/(\mathcal{m}\phi)$. Inserting this expression back into 
(\ref{ppaction2}) and writing $u^2$ explicitly, one finds
\be
S_{\rm p} = -\mathcal{m} \int d\lambda\,\phi \sqrt{g_{\mu\nu}\nd{x^\mu}{\lambda}
\nd{x^\nu}{\lambda}},
\label{ppaction3}
\ee
which is the usual expression for the action of a conformally coupled
massive particle (although (\ref{WGTppaction}) or
(\ref{ppaction2}) is immediately applicable also to massless particles
by setting $\mathcal{m}=0$, whereas (\ref{ppaction3}) is not).
Reparameterising the worldline in terms of $\xi$, such that $u^2=1$
throughout, from (\ref{ppaction3}) one again makes the identification
$d\tau = \phi\,d\xi$ for the particle proper time. In passing, we note
that this identification is also necessary in order for the trajectory
of a massive particle in conformal gravity not to depend on the
conformal frame in which it is calculated \cite{Hobson21}. 

Indeed, at
least in scale-invariant gravity theories, the above considerations
suggest that a compensator scalar field $\phi$ not only enables
dynamic generation of particle masses, but is also key in generating
a particle's (or observer's) proper time. More generally, the
necessity to introduce the scalar field into the calculation of
physical quantities renders them invariant under local scale
transformations. This means, for example, that one need not be
concerned about different physical properties of matter in different
parts of the universe as a consequence of a spatially-varying scalar
field, since such variations are automatically compensated for in
forming quantities that one can actually measure.  These ideas will be
developed further in a future publication.

In conclusion, we have revisited, clarified and extended our argument
in \cite{Hobson20} to confirm our original finding that the SCE does
{\it not} occur in WGTs, despite recent counterclaims in the
literature \cite{Quiros21}. In scale-invariant gravity theories, the
masses of the `ordinary' matter particles that make up atoms,
observers and clocks must be generated dynamically through the
interaction of the Dirac field with a scalar field $\phi$. By
determining the equations of motion for a particle in free fall, one
finds that neither the 4-velocity nor 4-momentum is parallel
transported along its worldline, but that the norm of the 4-velocity
is preserved, which addresses the original basis for suggesting the
existence of the SCE, in agreement with our findings in
\cite{Hobson20}. Moreover, while it is straightforward to verify the
observation in \cite{Quiros21} that the norm of a parallel-transported
vector of weight $w \neq -1$ depends on the path taken, this is not
relevant to the occurence of the SCE, since the 4-momentum of an atom
in free fall is transported in such a way that its norm is path independent,
thus verifying that the SCE does not occur. One may also extend
  our analysis to an ideal particle or clock that is not in free fall.
To avoid possible further confusion, we have also clarified and
  extended our
previous argument for defining an appropriately invariant proper time
variable along a particle worldline.

\vspace{-0.1cm}
\begin{acknowledgments}
\vspace*{-0.3cm}
We thank Israel Quiros for useful discussions.
\end{acknowledgments}

\end{document}